\documentclass[aps,prd,twocolumn,superscriptaddress,nofootinbib,
eqsecnum]{revtex4-1}
\usepackage{epsfig}
\usepackage{amsmath,amssymb}
\usepackage{graphicx}
\usepackage[mathcal]{eucal}
\usepackage{hyperref}

\newcommand{\eqn}[1]{Eq.~(\ref{#1})}
\newcommand{\eqns}[2]{Eqs.~(\ref{#1}),(\ref{#2})}
\newcommand{\xref}[1]{Ref.~\cite{#1}}

\newcommand{\secn}[1]{Sec.~\hspace{-2pt}\ref{#1}}

\def\plb{Phys.\ Lett.\ B}
\def\npb{Nucl.\ Phys.\ B}
\def\zpc{Z. Phys. C}

\def\us#1{\bf{#1}}

\def\npb#1({{\it Nucl.\ Phys.}\ $\us {B#1}$\ (}
\def\zpc#1({{\it Zeit.\ f\"ur Physik}\ $\us {C#1}$\ (}
\def\nuv#1({{\it Nuovo.\ Cim.}\ $\us {#1}$\ (}
\def\plb#1({{\it Phys.\ Lett.}\ $\us {B#1}$\ (}
\def\jlt#1({{\it JETP\ Lett.}\ $\us {#1}$\ (}
\def\cmp#1({{\it Comm.\ Math.\ Phys.}\ $\us {#1}$\ (}
\def\prp#1({{\it Phys.\ Rep.}\ $\us {#1}$\ (}
\def\prl#1({{\it Phys.\ Rev.\ Lett.}\ $\us {#1}$\ (}
\def\prd#1({{\it Phys.\ Rev.}\ $\us {D#1}$\ (}
\def\mplt#1({Mod.\ Phys.\ \Let.\ $\us {#1}$\ (}
\def\app#1({{\it Acta\ Phys.\ Pol.}\ $\us {#1}$\ (}

\newcommand{\betabar}{{\overline{\beta}}}
\def\wt#1{\widetilde{#1}}

\def\Btil{\tilde B}

\def\btil{\tilde b}

\def\be{\begin{equation}}
\def\ee{\end{equation}}
\def\bea{\begin{eqnarray}}
\def\eea{\end{eqnarray}}

\def\half{\frac{1}{2}}

\def\nn{\nonumber\\}
 
\def\gtil{\tilde g}
\def\pa{\partial}

\newcommand{\Tr}{{\rm{Tr}}}

\def\ie{{\it i.e., }}

\begin{document}


\title{Asymptotic freedom in certain $SO(N)$ and $SU(N)$ models.}

\author{Martin B. Einhorn$^{1,2*}$,
D.~R.~Timothy~Jones$^{1,3\dagger }$\\
$^{1}$Kavli Institute for Theoretical Physics, Kohn Hall,\\ 
University of California,
Santa Barbara, CA 93106-4030\\
$^{2}$Michigan Center for 
Theoretical Physics, 
University of Michigan,
Ann Arbor, MI 48109
$^{3}$Dept. of Mathematical Sciences,
University of Liverpool, Liverpool L69 3BX, U.K.
}\renewcommand{\thefootnote}{\fnsymbol{footnote}}
\footnotetext{meinhorn@umich.edu}
\footnotetext{drtj@liverpool.ac.uk}
\renewcommand{\thefootnote}{\arabic{footnote}}
\setcounter{footnote}{0}

\begin{abstract}
We calculate the $\beta$-functions for $SO(N)$ and $SU(N)$ gauge
theories coupled to adjoint and fundamental scalar representations,
correcting long-standing, previous results. We 
explore the constraints on $N$ resulting from requiring asymptotic
freedom for all couplings. When we take into account the actual allowed
behavior of the gauge coupling, the minimum value of $N$ in both cases
turns out to be larger than realized in earlier treatments. We also
show that in the large $N$ limit, both models have large regions of
parameter space corresponding to total asymptotic freedom.
\end{abstract}

\maketitle

\section{Introduction}\label{sec:intro}

The discovery of asymptotic freedom (AF) in
1973~\cite{Gross:1973id, Politzer:1973fx}\ heralded a new era in
particle physics. There was immediate interest in the extent to which AF
persists following the inclusion in a renormalizable gauge theory of
fermion and scalar multiplets. For fermions alone the question is easily
answered, but for scalars, or both fermions and scalars, it becomes
non-trivial. A pioneering and remarkably comprehensive analysis was
performed very early by Cheng {\it et al.}~(CEL)~\cite{Cheng:1973nv}. 
Under certain assumptions, a search for models of this type was carried 
out recently by Giudice {\it et al.}~\cite{Giudice:2014tma}, who labelled such 
models {\it Totally Asymptotically Free} (TAF). Other studies of this sort include 
Refs.~\cite{Holdom:2014hla, Pelaggi:2015kna}, who consider 
relativly low-scale ``unification" to a semi-simple group that is TAF.

Another important question arises once scalar multiplets are introduced,
being the nature and consequences of Spontaneous Symmetry Breaking (SSB)
in such AF theories; for example as to whether one can have an AF theory
with SSB to an abelian sub-group. CEL also address this issue, 
concluding that having enough scalar multiplets to achieve this is 
incompatible with AF. This explicit goal no longer seems essential; however a 
fully AF theory remains desirable.   

In a series of recent papers~\cite{Einhorn:2014gfa,Einhorn:2014bka,
Einhorn:2015lzy, Einhorn:2016mws}, we have addressed some other
aspects of these issues in the context of a gauge theory with 
scalar multiplets coupled to renormalizable, classically scale invariant
gravity. Our motivation in that work was twofold. Firstly, to
demonstrate examples of such theories that are AF and hence may be
termed Ultra-Violet (UV) complete; secondly, to show that in such
theories, SSB may occur via a variation on the perturbative Dimensional 
Transmutation mechanism first elucidated by Coleman and
Weinberg~\cite{Coleman:1973jx}.

Here we return to the AF issue, but in a class of theories with a more
complicated scalar sector than we have previously considered, namely two
distinct scalar representations transforming according to the adjoint
and the fundamental representations, with gauge groups $SO(N)$ and
$SU(N).$\footnote{The $SU(N)$ case for such scalars was considered 
by CEL, but we find some differences in our results both for the 
$\beta$-functions and for the minimum allowed value of $N.$} In contrast to 
Refs.~\cite{Giudice:2014tma, Holdom:2014hla, Pelaggi:2015kna}, we restrict 
our attention to grand unification in a simple group, even though this model is 
incomplete and does not contain the Standard Model (SM).

We assume the presence of a fermion sector contributing to the gauge
$\beta$-functions, but that concomitant Yukawa couplings are
sufficiently small that they are all asymptotically free.  As
usual~\cite{Cheng:1973nv}, they will then make negligible contributions
to the $\beta$-functions of  the quartic scalar couplings. We review the
flat space CEL calculations, where we find a number of significant
differences from their $\beta$-functions. In the light of these changes,
we reconsider the results for the minimum value of $N$ consistent with
AF in the case of both gauge groups. Here we find some differences from
previous results. For example, CEL correctly point out that the optimal
situation for AF of the scalar self couplings occurs for the {\it
minimum\/} of the (absolute) value of the gauge $\beta$-function
coefficient ($b_g$), which they choose to approximate by zero. However,
as we point out, this approximation can be inadequate to establish the
actual minimum value of $N$, and the {\it genuine\/} minimum of $b_g$
should be used in each case. This model for the  $SU(N)$ case has been
previously considered in  \xref{Buchbinder:1992rb}, with whose
$\beta$-functions we agree\footnote{The authors of
\xref{Buchbinder:1992rb}  did not mention their disagreements with CEL.
Our minimum value of $N$ differs from theirs.}.  We believe our
treatment of this $SO(N)$ model is new.

We gain further insight into the ``minimum value of $N$" issue by considering  
the large $N$ limit of these theories with appropriate rescaling of the scalar 
self-couplings. We shall discuss the extension of these results to 
renormalizable gravity elsewhere~\cite{EJ:largeN2}. 

The organization of the remainder of the paper is as follows: In 
Sections~\ref{sec:son} and \ref{sec:sun}, we give the beta-functions for the 
$SO(N)$ and $SU(N)$ models, respectively, and discuss the minimum value of 
$N$ consistent with TAF, comparing with earlier determinations. In 
\secn{sec:largeN}, we take up the large $N$ limits of these models and 
determine the ultraviolet stable FPs (UVFPs) for various associated 
fermionic content. After the Conclusions, \secn{sec:conclude}, we add 
two appendices deriving from the large $N$ models. In \secn{sec:analytic}, we 
indicate how the analytic solutions for the UVFPs can be obtained. In 
\secn{sec:CBZ}, we discuss the possible existence of an infrared 
fixed point (IRFP) for the gauge couplings at two-loops in certain cases.

\section{The $SO(N)$ Model}\label{sec:son}

The scalar potential of the theory is 
\bea\label{sonpot}
V(\Phi,\chi) &=& \frac{1}{2}\lambda_1 (\Tr\, \Phi^2)^2 
+ \lambda_2\Tr\, \Phi^4
+\frac{1}{8}\lambda_3 (\chi_i\chi_i)^2\nn &+&
\frac{1}{2}\lambda_4\chi_i\chi_i\Tr\, \Phi^2 
+\frac{1}{4}\lambda_5\chi_i \Phi_{ik}\Phi_{kj}\chi_j.
\eea
Here $\Phi = R^a\phi^a$, where $[a= 1,2 \ldots N(N-1)/2]$ represents 
a real adjoint representation, and $\chi_i\
[i=1,2\ldots N]$ is a real multiplet in the defining (fundamental)
representation, and $R^a$ are the associated antisymmetric $N\times N$
matrices normalised as usual so that 
\be
\Tr\, R^a R^b \equiv T(R)\delta^{ab},\quad\hbox{where}\quad T(R)= \frac{1}{2}. 
\ee
Thus, $\Tr[\Phi^2]=\phi^a\phi^a/2.$

Suppressing in each case a factor of $(16\pi^2)^{-1}$, the flat space 
$\beta$-functions are 
\bea
\beta_{g^2}
&=&-b_g (g^2)^2,\ b_g\equiv \frac{21N{-}43}{6}
-\frac{4}{3}T_F, \nn
\beta_{\lambda_1} &=& \Big(\frac{N(N{-}1)}{2}{+}8\Big)\lambda_1^2 
+2(2N{{-}}1)\lambda_1\lambda_2 +6\lambda_2^2\nn &+& N\lambda_4^2 
+ \lambda_4\lambda_5 - 6(N{-}2) g^2 \lambda_1 + 9g^4 ,\nn 
\beta_{\lambda_2} &=& (2N{-}1)\lambda_2^2 + 12\lambda_1\lambda_2+
\frac{1}{8}\lambda_5^2-6(N{-}2) g^2 \lambda_2\nn
&+& \frac{3(N{-}8)}{2}g^4,\nn 
\beta_{\lambda_3} &=& (N{+}8)\lambda_3^2{+}\frac{N(N{-}1)}{2} 
\lambda_4^2{+}\frac{N{-}1}{16}\lambda_5^2{+}
\frac{N{-}1}{2}\lambda_4\lambda_5\nn &-&
3(N{-}1) g^2 \lambda_3 +\frac{3(N{-}1)}{4}g^4,\nn
\beta_{\lambda_4} &=&4\lambda_4^2+\frac{1}{8}\lambda_5^2
+\lambda_5\bigg[\frac{N{-}1}{4}\lambda_1+\half\lambda_2+
+\half\lambda_3\bigg]\nn 
&+& x_4\bigg[\Big(\frac{N(N{-}1)}{2} +2\Big)\lambda_1+(2N{-}1)\lambda_2\nn
&+& (N{+}2)\lambda_3 - \frac{3(3N{-}5)}{2} g^2\bigg] +\frac{3}{2}g^4,\nn 
\beta_{\lambda_5} &=&\frac{N}{4}\lambda_5^2
+ \lambda_5\bigg[ 2\lambda_1+(N{-}1)\lambda_2
+2\lambda_3\nn &+&8\lambda_4
-\frac{3(3N{-}5) }{2}g^2\bigg] +3(N{-}4)g^4.
\label{sonbetas}
\eea
Here 
\be
 \Tr\, R_F^a R_F^b\equiv T_F\delta^{ab},
\ee
where the fermions transform according to the representation $R_F$, 
and the coefficient of $T_F$ in \eqn{sonbetas}\ reflects use of two-component 
or Majorana fermions.  
We obtained these results both by direct calculation and by use of the
RG equation for the effective potential (in the Landau gauge) in the
manner explained in the Standard Model context in
\xref{Einhorn:1982pp,Ford:1992pn}. 
The disagreements with CEL are 
in the coefficients of the following terms:
\begin{align}
\begin{split}
\beta_{\lambda_1} &{:}\, \lambda_4 \lambda_5;\ \  
\beta_{\lambda_2} {:}\, \lambda_5^2;\ \  
\beta_{\lambda_3} {:}\, \lambda_4 \lambda_5,\lambda_5^2;\ \nn 
\beta_{\lambda_4} &{:}\, \lambda_1\lambda_5,\lambda_2\lambda_5,
\lambda_3 \lambda_5,\lambda_5^2, g^4;\ \  
\beta_{\lambda_5} {:}\, \lambda_4 \lambda_5,\lambda_5^2, g^4.\!
\end{split}
\end{align}

To analyse the RG behavior of the couplings, it is convenient to 
introduce rescaled couplings $x_i = \lambda_i/g^2$, 
whereupon the ``reduced'' $\beta$-functions are:
\begin{align}
\betabar_{x_1} &{=} \Big(\frac{N(N{-}1){+}16}{2}\Big)x_1^2 
{+}6x_2^2{+}2(2N{-}1)x_1 x_2\nn
&{+}Nx_4^2{+}x_4x_5
{+}\big(b_g{-}6(N{-}2)\big)x_1 {+} 9,\!\nn 
\betabar_{x_2} &= (2N{-}1)x_2^2 +12x_1x_2+ \frac{1}{8} x_5^2
+ \big(b_g{-}6(N{-}2)\big)x_2\nn
&+\frac{3(N{-}8)}{2},\! \nn 
\betabar_{x_3} &= (N{+}8)x_3^2{+}\frac{N(N{-}1)}{2}x_4^2
{+}\frac{N{-}1}{16}x_5^2{+}\frac{N{-}1}{2}x_4x_5\nn
&{+}\big(b_g {-}3(N{-}1)\big)x_3 {+}
\frac{3(N{-}1)}{4},\! \nn 
\betabar_{x_4} &= 4x_4^2+\frac{1}{8}x_5^2
+x_5\bigg[\frac{N{-}1}{4}x_1+\half x_2 +\half x_3\bigg]+ \nn
&\hskip0.2in x_4\bigg[\Big(\frac{N(N{-}1)}{2}{+}2\Big)x_1+(2N{-}1)x_2+
 (N{+}2)x_3\nn &+ b_g-\frac{3(3N{-}5)}{2}\bigg] +\frac{3}{2}, \nn 
\betabar_{x_5} &= \frac{N}{4}x_5^2+x_5\bigg[2x_1+(N{-}1)x_2 
+2x_3 +8x_4\nn
&+b_g-\frac{3(3N{-}5)}{2}\bigg]+3(N{-}4).
\label{redSON}
\end{align}
In \eqn{redSON},
\be
\betabar_{x_i}\equiv\frac{dx_i}{du},\quad \hbox{where}\quad du\equiv  g^2(t)dt.
\ee

We now proceed to find and classify the Fixed Points (FPs) of this
system by setting all the reduced $\beta$-functions to zero. As long as
one has  $b_g > 0$, it is clear that any such FP (for finite $x_i$)
corresponds to TAF.  In fact, there are several FP solutions of this
system of equations but, it turns  out, only one is UV stable in all the
ratios $x_i.$  By UV stable, we mean that  the matrix
$S_{ij}\equiv\pa\overline{\beta}_{x_i}/\pa x_j$ has only negative 
eigenvalues at the FP, so that all ratios $x_i$  flow toward the FP
asymptotically.  We shall refer to such a point as a UVFP, even
though the original couplings are all TAF.

If any of the eigenvalues is zero, then one would have to go beyond the 
{\it linear} approximation to determine whether the associated flat
direction is in  fact a minimum. Should that test fail, one would have
to go beyond the  {\it one-loop} approximation unless one can identify
an exact symmetry ensuring  that such a flat direction persists to all
orders in perturbation theory. (It turns out  in the models considered
in this paper, such flat directions do not arise, so this  issue is
moot.)  For flat directions, there may also be non-perturbative effects 
such as instantons that lift the degeneracy but which we have not
investigated  presently.

For $SO(N)$, there will be a minimum value of $N$ consistent with the 
existence of a UVFP, and this minimum value of $N$ is generically a 
monotonically increasing function of $b_g$. For this reason, CEL set $b_g=0$ 
in order to obtain the minimum of $N$ consistent with a UVFP. However, this
reasoning results in incorrect results when we consider that, in fact, $b_g$ 
changes by discrete finite steps obtained by varying the fermion representations 
of the model.

If we assume a fermion content consisting of $n_F$ fundamental 
($N$-dimensional) two-component (or Majorana) representations, then 
\be
b_g= \frac{21N-43-4n_F}{6}.
\ee
Note that for AF we require $N > 2$. 
The minimum values of $b_g$ are obtained by taking $n_F$ as large as 
possible consistent with $b_g>0.$ These minima, $b_g^{\rm min}$, are shown 
in Table~\ref{SONbg}. 

Note that in the case $N= 3$ (mod 4), it in fact {\it is\/} possible to
have $b_g = 0$. However, in that case the two-loop correction to
$\beta_g$ is necessarily positive in the absence of Yukawa
couplings~\cite{Jones:1974pg}\ (which we  have been ignoring throughout)
and therefore this case fails to  be AF. 
\begin{table}
\begin{center}
\begin{tabular}{|c|c|c|c|c|} \hline
& & & & \\
$N$ & 3 (mod 4) & 4 (mod 4) & 5 (mod 4) & 6 (mod 4)\\ 
& & & & \\
\hline
& & & & \\
$b_g^{\rm min}$& $\frac{2}{3}$ & $\frac{1}{6}$ & $\frac{1}{3}$ & $\frac{1}{2}$\\
& & & & \\ \hline
\end{tabular}
\caption{\label{SONbg}Minimum value of $b_g$ in the class of $SO(N)$ models.}
\end{center}
\end{table}

With $b_g=0$, the minimum value of $N$ such that a UVFP results
is $N=10$. However,  this is {\it not\/} sustained when the actual value $b_g = 1/2$ 
is used from Table~1. For $b_g \neq 0$, the minimum value of $N$ for a UVFP is 
$N=12$. With $N=12$ and $b_g = 1/6$, we then find such a FP with 
\bea 
x_1 &=& 0.262953,\ 
x_2 = 0.111668,\ 
x_3 = 0.376914,\nn 
x_4 &=& 0.104270,\  
x_5 = 0.581883.
\eea
 
\section{The $SU(N)$ Model}\label{sec:sun}

In this case we have the scalar potential 
\bea\label{sunpot}
V(\Phi,\chi) &= \frac{1}{2}\lambda_1 (\Tr\, \Phi^2)^2 
+ \lambda_2\Tr\, \Phi^4
+\frac{1}{2}\lambda_3 (\chi_i^{\dagger}\chi^i)^2\nn
&+ \lambda_4\chi_i^{\dagger}\chi^i\Tr\, \Phi^2 
+\lambda_5\chi^{\dagger}_i \Phi^i{}_k\Phi^k{}_{j}\chi^j. 
\eea
Again $\Phi = R^a\phi^a$, where now $a= 1,2 \ldots N^2 - 1$. $\chi^i\
[i=1,2 \ldots N]$ is now a {\it complex\/} multiplet in the defining (fundamental)
representation, and $R^a$ are no longer (all) antisymmetric; they are again 
normalized as usual so that 
\be
\Tr\, R^a R^b = \frac{1}{2}. 
\ee
Thus, $\Tr[\Phi^2]=\phi^a\phi^a/2.$

As indicated in our Introduction (Section~\ref{sec:intro}), this model was examined in Chapter~9 of \xref{Buchbinder:1992rb}, with
$\beta$-functions given in Eq.~(9.26) in a slightly different notation.
We have however checked that their flat-space results are agreement with
ours below\footnote{\xref{Buchbinder:1992rb}\ does in fact have an
error, presumably inadvertent, in their formula for  $\beta_{f_3},$ in
which the coefficient of $g^2f_3$ should be $3(3N^2-1)/N,$ the  same as
given in their formula for $\beta_{f_4}.$}.  Our gravitational
corrections differ from theirs, but we shall discuss these
elsewhere~\cite{EJ:largeN2}.

Comparing with the corresponding expression in CEL, on the face of it 
the definition of the $\lambda_4$ terms differ by a factor of $4$.
However, in comparing results for the $\beta$-functions, it seems clear
that CEL have used {\it our\/} definition above in the actual
calculations. Nevertheless, there still remain significant differences 
in the results. Ours are as follows:
\bea
\beta_{g^2}&=&-b_g (g^2)^2,\ b_g\equiv \frac{21N{-}1}{3}
-\frac{4}{3}T_F,\nn
\beta_{\lambda_1} &=& (N^2{+}7)\lambda_1^2 
{+}\frac{4(2N^2{-}3)}{N}\lambda_1\lambda_2
{+}\frac{12(N^2{+}3)}{N^2}\lambda_2^2\nn
&+& 2N\lambda_4^2+4\lambda_4\lambda_5
-12Ng^2\lambda_1 {+} 18g^4,\nn 
\beta_{\lambda_2} &=& \frac{4(N^2{-}9)}{N}\lambda_2^2 +12\lambda_1\lambda_2+ \lambda_5^2
-12Ng^2 \lambda_2+3Ng^4,\nn 
\beta_{\lambda_3} &=& 2(N{+}4)\lambda_3^2+(N^2{-}1)\lambda_4^2
+\frac{(N{-}1)(N^2{+}2N{-}2)}{2N^2}\lambda_5^2\nn
&+&\frac{2(N^2{-}1)}{N}\lambda_4\lambda_5
- \frac{6(N^2{-}1)}{N} g^2 \lambda_3\nn
&+& \frac{3(N{-}1)(N^2{+}2N{-}2)}{2N^2} g^4,\nn 
\beta_{\lambda_4} &=& 4\lambda_4^2
+\lambda_4\bigg[(N^2{+}1)\lambda_1+\frac{2(2N^2{-}3)}{N}\lambda_2
+2(N{+}1)\lambda_3\bigg]
\nn
&+&
\lambda_5^2 + \lambda_5\bigg[\frac{N^2{-}1}{N}\lambda_1+
\frac{2(N^2{+}3)}{N^2}\lambda_2+2\lambda_3\bigg]\nn
&-&\frac{3(3N^2{-}1)}{N} g^2 \lambda_4 +3g^4,\nn
\beta_{\lambda_5} &=&\frac{N^2{-}4}{N}\lambda_5^2
+ \lambda_5\bigg[2\lambda_1+\frac{2(N^2{-}6)}{N}\lambda_2
+2\lambda_3+8\lambda_4\nn
&-&\frac{3(3N^2{-}1)}{N} g^2\bigg] 
+3Ng^4.
\label{sunbetas}
\eea
Assuming, as indicated above, that CEL actually used our definition of 
$\lambda_4$, 
we disagree with them  only in 
the coefficients of the following terms:
\begin{align}
\beta_{\lambda_4} &: g^4; \qquad
\beta_{\lambda_5}: \lambda_4 \lambda_5, g^4.
\label{termsSUN}
\end{align}

As before, the form for $b_g$ above in \eqn{bgSUN} assumes that the 
fermions are  two-component (or Majorana).   For example, if we have an
arbitrary number $n_F$ of fermions in the  $N$-dimensional
representation, then $T_F=1/2$ and 
\be 
b_g =\frac{21N-1}{3}-\frac{2n_F}{3}. \label{bgSUN} 
\ee 
However the
$N$-dimensional representation of $SU(N)$  gives non-zero triangle
anomalies for $N \geq 3$, so, in that case,   $n_F$ above is necessarily
even.  Using \eqn{bgSUN}, the results for  $b_g^{\rm min}$ are shown in
Table~\ref{SUNbg}. (One can achieve $b_g =0$ in
the case $N = 5$ (mod 4), but we eschew this as before because of the
effect of two-loop corrections.)

The corresponding reduced $\beta$-functions ($x_i \equiv \lambda_i/g^2$) are:
\bea\label{reducedsunbeta}
\betabar_{x_1} &=& (N^2{+}7)x_1^2 {+}\frac{4(2N^2{-}3)}{N}x_1x_2
{+}\frac{12(N^2{+}3)}{N^2}x_2^2\nn &+& 2Nx_4^2
+ 4x_4x_5{+}
\big(b_g{-}12N\big) x_1 {+} 18,\nn 
\betabar_{x_2} &=& \frac{4(N^2{-}9)}{N}x_2^2 +12x_1x_2+ x_5^2
{+}\big(b_g{-}12N\big) x_2+3N, \nn 
\betabar_{x_3} &=& 2(N{+}4)x_3^2{+}(N^2{-}1)x_4^2+
\frac{(N{-}1)(N^2+2N{-}2)}{2N^2}x_5^2\nn
&+& \frac{2(N^2{-}1)}{N}x_4x_5+\nn
&~& \Big(b_g {-} \frac{6(N^2{-}1)}{N}\Big) x_3 + 
\frac{3(N{-}1)(N^2{+}2N{-}2)}{2N^2}, \nn 
\betabar_{x_4} &=& 4x_4^2+ x_4\bigg[(N^2{+}1)x_1+\frac{2(2N^2{-}3)}{N}x_2
+2(N{+}1)x_3\bigg]\nn 
&+&x_5^2+x_5\bigg[\frac{N^2{-}1}{N}x_1+\frac{2(N^2{+}3)}{N^2}x_2+2x_3\bigg]\nn
&+&\Big(b_g {-}\frac{3(3N^2{-}1)}{N}\Big) x_4 +3 ,\nn 
\betabar_{x_5} &=& \frac{N^2{-}4}{N}x_5^2+x_5\bigg[
2x_1+\frac{2(N^2{-}6)}{N}x_2
+2x_3+8x_4\nn
&+&\Big(b_g {-}\frac{3(3N^2{-}1)}{N}\Big) \bigg]+3N. 
\eea

\begin{table}
\begin{center}
\begin{tabular}{|c|c|c|c|c|} \hline
& & & & \\
$N$ & 2 (mod 4) & 3 (mod 4) & 4 (mod 4) & 5 (mod 4)\\ 
& & & & \\
\hline
& & & & \\
$b_g^{\rm min}$& $\frac{1}{3}$ & $\frac{2}{3}$ & $1$ & $\frac{4}{3}$\\
& & & & \\ \hline
\end{tabular}
\caption{\label{SUNbg}Minimum value of $b_g$ in the class of $SU(N)$ models.}
\end{center}
\end{table}

For this model, using the approximation $b_g=0,$ the smallest value
of $N$ required to have all couplings AF was given as $N_{min}=7$ in
\xref{Cheng:1973nv}, using incorrect $\beta$-functions, and as
$N_{min}=8$ in \xref{Buchbinder:1992rb}, using the same
$\beta$-functions as ours. For $N=8,$ the actual minimum value is
$b_g^{min}=1,$ for which we find the model is \emph{not} AF.  For $N=9,$
we have $b_g^{min}=4/3,$ for which the model is AF with its UVFP at
\bea
x_1 &=& 0.386000,\ x_2 = 0.293121,\ x_3 = 0.502429,\nn 
x_4 &=& 0.195158,\ x_5 = \
0.398832.
\eea

\section{The Large $N$ limit}\label{sec:largeN}

Let us consider the large $N$ limit of this class of theories. Of
course, as shown many years ago by 't~Hooft~\cite{tHooft:1973alw} 
for $SU(N)$, the relevant graphs in the large $N$ limit 
are {\it planar}\/; summing these graphs to obtain the full 
leading $N$ approximation 
has proved elusive, even for the pure Yang-Mills theory,
and despite the fact that there must exist a classical 
Master Equation~\cite{Witten:1979pi}.
Consequently, to salvage perturbative believability, our results will 
still require the relevant couplings to be small. 
Nevertheless, the results have features of interest. 

Let us begin by considering the $SU(N)$ case. (The results for $SO(N)$ turn 
out to be essentially the same and will be given below.) Because the
gauge contribution to $b_g$ naturally grows as $N$,  $\wt{b}_g\equiv
b_g/N$ remains finite as $N\to\infty.$ Then, as 't~Hooft 
showed~\cite{tHooft:1973alw}, defining a rescaled gauge coupling  
$\wt{g}{\,}^2\equiv N g^2,$ its $\beta$-function satisfies 
\be
\beta_{\gtil^2}=-\wt{b}_g(\gtil^2)^2
\ee
Thus, in the limit $N{\to}\infty$, $g{\to}0$ for fixed $\gtil^2,$
$\beta_{\tilde{g}^2}$  remains finite. Similarly, if we rescale the couplings
$\lambda_i$ in a certain way,  the resulting $\beta_{\wt{\lambda}_i}$
will have finite limits in terms of rescaled  couplings
$\wt{\lambda}_i.$ This requires  
\bea
\lambda_1
&=& {\wt{\lambda}_1}/{N^2}, \lambda_2={\wt{\lambda}_2}/{N},
\lambda_3={\wt{\lambda}_3}/{N}, \lambda_4={\wt{\lambda}_4}/{N^{p_4}},\nn
\lambda_5 &=& {\wt{\lambda}_5}/{N},
\label{eq:rescaling}
\eea
for $3/2\leq p_4\leq 2.$  This ambiguity in the rescaling of $\lambda_4$ reflects a nonuniformity of the limiting behavior. 
For $3/2<p_4<2,$  all dependence on $\wt{\lambda}_4$ drops out except in 
$\beta_{\wt{\lambda}_4},$ and we find
\bea\label{sunbetap4}
\beta_{\wt{\lambda}_1}&=&\wt{\lambda}_1^2+
8\wt{\lambda}_1\wt{\lambda}_2+
12\wt{\lambda}_2^2+
18\tilde{g}^4{-}12\tilde{g}^2\,\wt{\lambda}_1,\nn
\beta_{\wt{\lambda}_2}&=&4\wt{\lambda}_2^2+
3\tilde{g}^4 {-}12\tilde{g}^2\,\wt{\lambda}_ 2 ,\nn
\beta_{\wt{\lambda}_3}&=&2\wt{\lambda}_3^2{+}
\frac{1}{2}\wt{\lambda}_5^2{+}
\frac{3}{2}\tilde{g}^4{-}6\tilde{g}^2\,\wt{\lambda}_3,\\
\beta_{\wt{\lambda}_4}&=&\wt{\lambda}_4\left(
\wt{\lambda}_1{+}
4\wt{\lambda}_2{+} 
2\wt{\lambda}_3
{-}9\tilde{g}^2\right),\nn
\beta_{\wt{\lambda}_5}&=&
2\wt{\lambda}_2\,\wt{\lambda}_5{+}\wt{\lambda}_5^2\!+
3\tilde{g}^4{-}9 \tilde{g}^2\,\wt{\lambda}_ 5.\nonumber
\eea
Inasmuch as $\beta_{\wt{\lambda}_4}$ is linear in $\wt{\lambda}_4,$ it 
differs from the others and from the finite $N,$ \eqn{sunbetas}, $\beta$-functions.  Consequently, it has a FP at $\lambda_4=0,$ independent of the values of the other couplings.  It turns out that, when one forms the reduced $\beta$-functions in terms of the ratios $\wt{y}_i\equiv \wt{\lambda}_i/\tilde{g}^2,$ $y_4=0$ is in fact a UVFP for \eqn{sunbetap4}.

At the extreme values, $p_4=3/2$ or $p_4=2,$ other terms survive.  
In the case, $p_4=3/2,$ there are quadratic terms in $\wt{\lambda}_4$ that survive in $\beta_{\wt{\lambda}_1}$ and $\beta_{\wt{\lambda}_3}$, to wit,
\begin{align}
\begin{split}
\beta_{\wt{\lambda}_1}&=\wt{\lambda}_1^2+
8\wt{\lambda}_1\wt{\lambda}_2{+}
12\wt{\lambda}_2^2{+}2\wt{\lambda}_4^2{+}
18\tilde{g}^4{-}12\tilde{g}^2\,\wt{\lambda}_1,\\
\beta_{\wt{\lambda}_3}&=2\wt{\lambda}_3^2{+}
\frac{1}{2}\wt{\lambda}_5^2{+}\wt{\lambda}_4^2 {+}
\frac{3}{2}\tilde{g}^4{-}6\tilde{g}^2\,\wt{\lambda}_3.
\end{split}
\end{align}
The remaining three $\beta$-functions are the same as in \eqn{sunbetap4}.  
It turns out that the UVFP remains at $\lambda_4=0$ in this case, so the presence of these additional terms does not change the values of the UVFP from the case $3/2<p_4<2,$ \eqn{sunbetap4}.  They will however affect the running of the couplings away from the FP.

For $p_4=2,$ all $\beta_{\wt{\lambda}_i}$ for $i\neq4,$ are unchanged,
whereas $\beta_{\wt{\lambda}_4}$ becomes
\bea\label{sunbetat2}
\beta_{\wt{\lambda}_4} &=& \wt{\lambda}_5\left(\wt{\lambda}_1+
2\wt{\lambda}_2+ 2\wt{\lambda}_3
\right)+\wt{\lambda}_5^2+3\tilde{g}^4\nn &+&
\wt{\lambda}_4\left(\wt{\lambda}_1+4\wt{\lambda}_2+ 
2\wt{\lambda}_3{-}9\tilde{g}^2\right)\!.
\eea
In fact, this, together with the other $\beta$-functions from \eqn{sunbetap4}, 
are an excellent approximation to the large-$N$ behavior of the exact equations, \eqn{sunbetas}. The other choices for $p_4$ do not appear to be physically relevant.

For $p_4=2,$ the reduced $\beta$-functions in terms of 
$\wt{y}_i\equiv \wt{\lambda}_i/\tilde{g}^2,$ are
\vskip-0.25in
\bea\label{betabartildesuNflat2}
\betabar_{\wt{y}_1}&=&\wt{y}_1^2+
12\wt{y}_2^2+18-(12-\wt{b}_g-8\wt{y}_2)\wt{y}_1,\nn
\betabar_{\wt{y}_2}&=&4\wt{y}_2^2
+3 -(12-\wt{b}_g)\wt{y}_ 2 ,\nn
\betabar_{\wt{y}_3}&=&2\wt{y}_3^2+
\frac{1}{2}\wt{y}_5^2 +
\frac{3}{2}-(6-\wt{b}_g)\,\wt{y}_ 3,\nn
\betabar_{\wt{y}_4}&=&\wt{y}_5\big(\wt{y}_1+
2\wt{y}_2+2\wt{y}_3\big)+\wt{y}_5^2+3\nn
&+& \wt{y}_4\big(
\wt{y}_1+4\wt{y}_2+2\wt{y}_3
-(9-\wt{b}_g)\big),\nn
\betabar_{\wt{y}_5}&=&
\wt{y}_5^2\!+
3-(9-\wt{b}_g-2\wt{y}_2)\wt{y}_ 5.
\eea
\vskip-0.2in

\begin{table}[hbt]
\begin{center}
\begin{tabular}{|c|c|c|c|c|c|} \hline
& & & & & \\
$\wt{b}_g$ & $\wt{y}_1$  & $\wt{y}_2$ & $\wt{y}_3$ & $\wt{y}_4$ & $\wt{y}_5$ \\ 
& & & & & \\
\hline
0.& 2.64270& 0.275255 & 0.289413  & 0.970346 &0.371374\\
\hline
$1/3$& 2.94605& 0.284989 & 0.312552 & 1.20422 &0.389234\\
\hline
1/2& 3.15683 & 0.290153 & 0.325788  & 1.39047 &0.398894\\
\hline
3/4&  3.67495& 0.298306 & 0.348280  & 1.94791 &0.414424\\
\hline
0.84798& 4.36728& 0.301646 & 0.358128  & 2.99190 &0.420885\\
\hline
\end{tabular}
\caption{\label{suUVFPs} UVFPs for $SU(\infty).$}
\end{center}
\end{table}

Solving simultaneously the equations 
$\betabar_{\wt{y}_i}=0,$ we find several FPs, one of which is UV stable.
The values of this UVFP for various values of $\wt{b}_g$ are given in 
Table~\ref{suUVFPs}. For $\wt{b}_g\gtrsim 0.84798,$ there are no real FPs.

The results for $SO(N)$ are precisely analogous to those above.  With the 
definitions of the self-couplings $\lambda_i$ in \eqn{sonpot}, the 
$\beta$-functions 
for the gauge and self-couplings are
\bea\label{eq:betaflatN}
\beta_{\tilde{g}^2}&=&-\wt{b}_g \tilde{g}^4,\nn
\beta_{\wt{\lambda}_1}&=&\frac{1}{2}\wt{\lambda}_1^2+6\wt{\lambda}_2^2
+9\tilde{g}^4+\big(4\wt{\lambda}_2-6\tilde{g}^2\big)\wt{\lambda}_1,\nn
\beta_{\wt{\lambda}_2}&=& 
2\wt{\lambda}_2^2+\frac{3}{2}\tilde{g}^4{-}6\tilde{g}^2\wt{\lambda}_2,\nn
\beta_{\wt{\lambda}_3}&=&\wt{\lambda}_3^2{+}
\frac{1}{16}\wt{\lambda}_5^2
+\frac{3}{4}\tilde{g}^4{-}3\tilde{g}^2\wt{\lambda}_3,\\
\beta_{\wt{\lambda}_4}&=& \wt{\lambda}_5\bigg[ \frac{1}{4}\wt{\lambda}_1+
\frac{1}{2}\wt{\lambda}_2+\frac{1}{2}\wt{\lambda}_3\bigg]
+\frac{1}{8}\wt{\lambda}^2_5+\frac{3}{2}\tilde{g}^4\nn &+&
\wt{\lambda}_4\bigg[\frac{1}{2}\wt{\lambda}_1+ 2\wt{\lambda}_2+
\wt{\lambda}_3-\frac{9}{2}\tilde{g}^2\bigg],\nn
\beta_{\wt{\lambda}_5}&=&\frac{1}{4}\wt{\lambda}_5^2+3\tilde{g}^4+
\Big(\wt{\lambda}_2-\frac{9}{2}\tilde{g}^2\Big)\wt{\lambda}_5.\nonumber
\eea
The $\tilde{\lambda}_i$  above are defined as in Eq. 4.2, with $p_4 =2.$

Defining once again, $y_i\equiv\wt{\lambda}_i/\tilde{g}^2,$ the reduced $\beta$-functions are 
\bea
\label{betabartildesoNflat2}
\betabar_{\wt{y}_1}&=&\frac{1}{2}\wt{y}_1^2+6\wt{y}_2^2+9
+\big(4\wt{y}_2+\wt{b}_g-6\big)\wt{y}_1,\nn
\betabar_{\wt{y}_2}&=& 
2\wt{y}_2^2+\frac{3}{2}+(\wt{b}_g-6)\wt{y}_2,\nn
\betabar_{\wt{y}_3}&=&\wt{y}_3^2{+}
\frac{1}{16}\wt{y}_5^2
+\frac{3}{4}+(\wt{b}_g{-}3)\wt{y}_3,\nn
\betabar_{\wt{y}_4}&=& \wt{y}_5\bigg[ \frac{1}{4}\wt{y}_1+
\frac{1}{2}\wt{y}_2+\frac{1}{2}\wt{y}_3\bigg]
+\frac{1}{8}\wt{y}^2_5+\frac{3}{2}\nn 
&+&
\wt{y}_4\bigg[\frac{1}{2}\wt{y}_1+ 2\wt{y}_2+
\wt{y}_3+\wt{b}_g-\frac{9}{2}\bigg],\nn
\betabar_{\wt{y}_5}&=&\frac{1}{4}\wt{y}_5^2+3+
\wt{y}_5\Big(\wt{y}_2+\wt{b}_g-\frac{9}{2}\Big).
\eea
As with $SU(N),$ we find several FPs, of which one is UV stable. The
values of this UVFP for various values of $\wt{b}_g$ are given in 
Table~\ref{soUVFPs}. For $\wt{b}_g\gtrsim 0.42399,$ there are no real
FPs.  

\begin{table}[hbt]
\begin{center}
\begin{tabular}{|c|c|c|c|c|c|} \hline
& & & & & \\
$\wt{b}_g$ & $\wt{y}_1$  & $\wt{y}_2$ & $\wt{y}_3$ & $\wt{y}_4$ & $\wt{y}_5$ \\ 
& & & & & \\
\hline
0.& 2.64270& 0.284989 & 0.310944  & 1.17978 & 0.710102\\
\hline
1/6& 2.94605& 0.290153 & 0.325788  & 1.39047 & 0.741044\\
\hline
$1/3$& 3.45350& 0.295531& 0.338224 & 1.64814 & 0.774966\\
\hline
5/12&  4.08657 &  0.301141 & 0.354074 & 2.44429 & 0.793191\\
\hline
0.42399& 4.36728& 0.301646 & 0.355550  & 2.90078 & 0.794836\\
\hline
\end{tabular}
\caption{\label{soUVFPs} UVFPs for $SO(\infty).$}
\end{center}
\end{table}

A cursory comparison of Tables~\ref{suUVFPs} and \ref{soUVFPs} 
indicates that many of the rows for the UVFP $\wt{y}_n$ are approximately the 
same provided, in Table~\ref{soUVFPs}, one doubles $\wt{b}_g$ and halves 
$\wt{y}_5.$  Most entries then agree at least in their first two significant figures!  This 
comes about because the leading term in $b_g$ is proportional to $C(G),$ 
which, for $SO(N),$ is $N/2,$ half that of $SU(N).$ To 
understand the factor of two in $\wt{y}_5,$ we must compare the the 
normalization of $\lambda_5$ in the potentials, \eqns{sonpot}{sunpot}.  
Recalling that $\chi_i$ is complex for $SU(N)$ and real for $SO(N),$ we would 
anticipate the couplings might correspond at large $N$ if  
$\lambda_5$ were replaced by $\lambda_5/2$ in the potential for $SU(N).$ 

On the other hand, if, as with $SO(10)$, one were to add a fermion in the 
smallest spinor representation of $SO(2n),$ for which $T(R)=2^{(n{-}4)},$ then 
obviously the condition that $b_g>0$ will be violated at some finite $n.$  (In fact, 
one must have $n\leq10.$) Thus, there would be no large-$N$ scaling limit in 
such a case.

The equations \eqns{betabartildesuNflat2}{betabartildesoNflat2} are 
sufficiently simple to be solvable analytically (as functions of
$\wt{b}_g$) for the FPs of the $\beta$-functions, in particular, for the
UVFP. This is described in Appendix~\ref{sec:analytic}. In practice, it
is actually easier simply to  solve for the FPs numerically.  Knowing
from the preceding which of the FPs is the candidate UVFP, one can
easily check whether the eigenvalues of the stability matrix $S_{ij}$
are all negative.  In fact, since the UVFP occurs for positive
$\wt{y}_i,$ we can be confident that it is 
unique~\cite{Bais:1978fv}\footnote{The example given in
\xref{Bais:1978fv}\ unfortunately uses the $\beta$-functions of
\xref{Cheng:1973nv} for the model we have treated here. As we have
stated, some of those $\beta$-functions are incorrect, but, in their 
application, the qualitative conclusions of \xref{Bais:1978fv} remain unchanged.}.

With reference to the first rows of Tables~\ref{suUVFPs} and 
\ref{soUVFPs}, it is clear that for large but finite $N,$ $\btil_g$ is very 
small. One ought to wonder whether the two-loop corrections to the 
$\beta$-functions might not be equally large in certain cases. Such a 
possibility has been examined in the 
past~\cite{Caswell:1974gg, Banks:1981nn} and leads to the idea that 
there may be a finite IRFP in $g^2,$ a so-called CBZ FP. We elaborate 
on this possibility in Appendix~\ref{sec:CBZ}.

\section{Conclusions}\label{sec:conclude}

We have presented the flat space one-loop $\beta$-functions for both
$SU(N)$ and $SO(N)$  gauge theories coupled to scalar multiplets in both
the adjoint and fundamental representations. Both cases were originally
studied in CEL; our results differ from theirs in a number of terms, as
do our conclusions regarding the minimum  values of $N$ consistent with
TAF, \ie asymptotic freedom of all the couplings. In the $SU(N)$ case,
our results for the $\beta$-functions agree with those presented in BOS 
(though not so, as we shall discuss elsewhere~\cite{EJ:largeN2},   when
extended to renormalizable gravity). Instead of simply approximating the
 minimum allowed value of $b_g>0$ by zero, we paid particular attention
to the  actual minimum for an essentially arbitrary choice of
fermion representations (Tables~\ref{SONbg} \& \ref{SUNbg}), except for
spinor representations, for which there is no large $N$ scaling limit
that is still TAF.

One interesting result in the case of  $SO(N)$ is that the smallest
allowed value of $N$ is greater than  $N=10$ (as it is  for $b_g=0$)
when the actual $b^{\mathrm{min}}_g=1/2.$ The minimum may go  even
higher than $N=12$ when additional scalars are included in order to have
 appropriate Yukawa couplings to accommodate the SM fermion spectrum 
and to incorporate electroweak symmetry breaking.

For $SU(N),$ we found that the smallest value of $N$ for which all
couplings are AF is $N_{min}=9,$ for which $b^{\mathrm{min}}_g=4/3.$ 
This is to be compared with $N_{min}=7$ in \xref{Cheng:1973nv},
using incorrect $\beta$-functions, and $N_{min}=8$ in
\xref{Buchbinder:1992rb}, using correct $\beta$-functions but
taking $b_g=0.$

We also discussed the large $N$ limit in both theories,  with couplings
appropriately rescaled so as to render the $\beta$-function coefficients
finite. One result there is that there is an allowed \emph{maximum}
value of $b_g$ for large $N$ beyond which there is no real UVFP. It
is about $0.85N$ for $SO(N)$ and $0.42N$ for $SU(N),$ so the allowed
range of choices for the fermion representations is not nearly so
restrictive as suggested by choosing $N$ to be as small as permitted,
and it may become much easier to accommodate the three generations of
fermions in the SM. These  results are, we believe, novel and
interesting.

These calculations constitute part of our efforts to develop a UV
complete, TAF theory coupled to renormalizable, scale-invariant  
gravity that is realistic, i.e., one that leads to the  Standard Model
plus Einstein-Hilbert gravity at low energies. We plan to extend our
results here to  incorporate gravitational couplings and to explore
whether  Dimensional Transmutation can generate both gauge symmetry
breaking and a Planck mass term, along the lines
of~\xref{Einhorn:2016mws}.  Then, for a realistic model, other scalar
representations and the effect  of Yukawa couplings must be considered.
We showed in ~\xref{Einhorn:2016mws} how breaking of $SO(10)$ to
$SU(5)\times U(1)$ can occur in a scale invariant model;  one
outstanding problem is how further breaking may be engineered,
eventually to the Standard Model Gauge group. The results in this paper
suggest that it will require $N_{min}\geq 12$ for  $SO(N)$ and
$N_{min}\geq 9$ for $SU(N),$ and these minimum values may be even larger
after adding additional scalars needed to account for fermion masses and
 to break down to the SM gauge symmetries. Renormalizable gravity
makes relatively small changes to the flat space results near the UVFP,
but there remains the issue of unitarity in such theories.

\begin{acknowledgments}
DRTJ thanks  KITP (Santa Barbara), where part of
this work was done, for hospitality. This research was supported in part
by the National Science Foundation under Grant 
No.\ NSF \nolinebreak PHY11-25915
 and by the Baggs bequest.
 \end{acknowledgments}
 
\begin{appendix}

\section{Analytic solutions for the large-$N$ fixed points}
\label{sec:analytic}

As mentioned in the text, 
\eqns{betabartildesuNflat2}{betabartildesoNflat2} are sufficiently
simple that, given $\wt{b}_N,$ their FPs can be
analytically determined.  Although all FPs may be so determined, we
shall focus on finding the UVFP.

Consider first the $SU(N)$ case, \eqn{betabartildesuNflat2}. Note 
that $\betabar_{\wt{y}_2}$ is a function of $\wt{y}_2$ only. 
It will have real zeros if and only if the discriminant of the quadratic
is positive:

\be
(6-\wt{b}_g/2)^2-12>0.
\ee
Assuming $12-\wt{b}_g>0,$ then the two FPs occur for $\wt{y}_2>0,$ and it is 
easy to see that the smaller is the UVFP.  We can input this value of 
$\wt{y}_2$ into the other four $\beta$-functions to search for a UVFP in the other 
$\wt{y}_n.$  This enables us to solve explictly for the FPs in $\wt{y}_1$ from 
$\betabar_{\wt{y}_1}=0$ and for $\wt{y}_5$ from $\betabar_{\wt{y}_5}=0,$ along 
with further constraints on $\wt{b}_g$ arising from requiring the equations to 
have real roots. 
In each case, we can choose the root of the quadratic equation having negative 
slope for fixed values of the other $\wt{y}_n,$ giving us further candidates for 
the UVFP.  Given $\wt{y}_5$, we can then solve for $\wt{y}_3$ from 
$\betabar_{\wt{y}_3}=0,$ and choose the smaller root once again.  
So now we have candidate values for $\wt{y}_k, \{k=1,2,3,5\}.$
Finally, $\betabar_{\wt{y}_4}$ is linear in $\wt{y}_4,$ so it has a unique root that 
can be expressed in terms of the solutions for the other $\wt{y}_n.$  In principle, 
it could be positive or negative, but it is a UVFP only if the coefficient is 
negative, i.e., only for 
\be
\wt{y}_1+4\wt{y}_2+2\wt{y}_3<9-\wt{b}_g.
\ee
Thus, the root for $\wt{y}_4$ is also positive.
Since each of the UVFPs is known as a function of $\wt{b}_g,$ this inequality
may further restrict the range of $\wt{b}_g$ within which there are real solutions 
for all the UVFPs. (See Table~\ref{suUVFPs}.)

Thus we arrive at a unique candidate for the UVFP, within a restricted
range of  $\wt{b}_g.$  We cannot immediately conclude that this is a
UVFP because the  stability matrix  $S_{m,n}\equiv
\pa\betabar_{\wt{y}_m}/\pa\wt{y}_n$ at a FP has non-zero off-diagonal 
terms (except in the case of $\betabar_{\wt{y}_2}.)$ In the preceding, we
only took into account the signs of the diagonal entries in each case.
One must verify that the true eigenvalues at the putative UVFP have the
signs of the diagonal entries.  In fact, they do.  

The solution for the $SO(N)$ case, \eqn{betabartildesoNflat2} can be obtained 
in precisely the same manner.  The only changes are in the numerical coefficients 
of the couplings.

\section{The CBZ Infrared Fixed Point}\label{sec:CBZ}

While $b_g >0$ is required for AF of the gauge coupling, to obtain  AF
for the quartic scalar couplings as well it is optimal to employ the
smallest possible value of $b_g$. This suggests the possible existence
of a CBZ~\cite{Caswell:1974gg, Banks:1981nn} infra-red stable fixed
point (IRFP);  in other words, the basin of attraction of the UVFP at
$g^2=0$ is finite\footnote{We thank a referee for a suggestion that 
inspired the following remarks.}.  Writing  
\be \beta_{g^2} =
-\frac{b_g}{16\pi^2} g^4 + 2 \frac{B}{(16\pi^2)^2}g^6, 
\ee
we have in general (in the absence of Yukawa couplings) that
\be
b_g = 2\bigg(\frac{11}{3}C_G - \frac{2}{3}T_F - \frac{1}{6}T_S\bigg)
\ee
and
\bea
B  &=&\frac{10}{3}C_G T_F +2\sum C_{F_\alpha}T_{F_\alpha} 
+ \nn &~&2\sum C_{S_\beta}T_{S_\beta}+\frac{1}{3}C_G T_{S}-\frac{34}{3} C_G^2.
\eea
Here $T_F = \sum T_{F_\alpha}$ and $T_S = \sum T_{S_\beta}$ where we
label the irreducible fermion and scalar representations by
$\alpha,\beta$ respectively.

It was first noted by Caswell~\cite{Caswell:1974gg} that, in a gauge 
theory with fermions (but no scalars), 
for $b_g = 0$, $B > 0$.  It
follows that for $b_g > 0$ but sufficiently small, there exists a
perturbatively believable IRFP corresponding to
\be
\frac{g^2_{IR}}{16\pi^2} = \frac{b_g}{2B}.
\ee
In the case of a gauge theory with scalars (but no fermions) or with
both scalars and fermions the corresponding result is less obvious, but
a detailed examination of the possible quadratic Casimir operators confirms
that the same result holds in these cases, too~\cite{Bond:2016dvk}.

Given the proximity of the IRFP to the origin, it is clear that there is
only a limited range of  values, $0 < g < g_{IR}$, of $g$ at some
reference scale (the GUT scale for instance),  corresponding to AF. For
$g>g_{IR},$ then $g$ approaches a Landau pole in the UV, {\it i.e.,} 
perturbation theory breaks down. In particular:
at large $N$, for either $SO(N)$ or $SU(N),$ it is easy to see that 
$B{\to} kN^2,$ where $k$ is a constant.  In the large $N$ limit, 
we define $\btil_g \equiv b_g/N,$ $\Btil \equiv B/N^2,$ and 
$\gtil\,^2 \equiv Ng^2,$ as in \secn{sec:largeN}. Then 
\be
\frac{\wt{g}_{IR}^{\;2}}{16\pi^2} = \frac{\wt{b}_g}{2\wt{B}}.
\ee
It is thus clear that for very small $\btil_g$, corresponding to the
first rows of Tables~\ref{suUVFPs} and \ref{soUVFPs}, the range of 
$\gtil$ corresponding to AF is actually very limited. This may constrain 
model building involving renormalizable quantum gravity of the kind 
envisaged in \xref{Einhorn:2016mws}, where it was important  that the 
region of coupling constant space corresponding to Dimensional 
Transmutation and spontaneous symmetry breaking lay within the 
basin of attraction of the UVFP of coupling constant ratios 
corresponding to AF of all couplings.
Conversely, should the IRFP of the gauge coupling be approached 
in the IR, the resulting theory  would probably become strongly coupled, 
because the gravitational self-couplings increase in the IR. 
Then we would expect a QCD-type phase transition before the 
gauge coupling reaches its IRFP, unless all the other couplings also displayed CBZ behaviour in the IR limit.

\end{appendix}
\vfill
\pagebreak

\end{document}